\documentclass[aps,12pt,preprint]{revtex4}
\usepackage{graphicx}
\usepackage{amsmath}
\usepackage{hyperref}
\usepackage{bm}
\usepackage[dvips]{color}

\newcommand{\etal} {\textit{et al.}}

\newcommand{\fzd} {Institute of Ion Beam Physics and Materials Research, Forschungszentrum Dresden-Rossendorf, 01314 Dresden, Germany}

\newcommand{\mg} {Mn$_5$Ge$_3$}

\begin{document}

\title{Memory effect of Mn$_5$Ge$_3$ nanomagnets embedded inside a Mn-diluted Ge matrix }

\author{Shengqiang~Zhou}
\email[Electronic address: ]{S.Zhou@fzd.de} \affiliation{\fzd}
\author{Artem~Shalimov}
\author{Kay~Potzger}
\author{Nicole~M.~Jeutter}
\author{Carsten~Baehtz}
\author{Manfred~Helm}
\author{J\"{u}rgen~Fassbender}
\author{Heidemarie~Schmidt}
\affiliation{\fzd}

\begin{abstract}
Crystalline \mg~nanomagnets are formed inside a Mn-diluted Ge matrix using Mn ion implantation. A temperature-dependent memory effect and slow
magnetic relaxation are observed below the superparamagnetic blocking temperature of \mg. Our findings corroborate that the observed
spin-glass-like features are caused by the size distribution of \mg~nanomagnets, rather than by the inter-particle interaction through the
Mn-diluted Ge matrix.
\end{abstract}
\maketitle

It was recently pointed out that the
integration of \mg~within the Ge matrix is indeed quite promising for spin injection in a silicon-compatible geometry \cite{zeng:5002}. \mg~is a
ferromagnet with a Curie temperature (\emph{T}$_C$) of 296 K and with a large spin polarization \cite{PhysRevB.70.235205}. Therefore,
considerable work has been done in fabricating \mg~epitaxial films \cite{zeng:5002,PhysRevB.70.205340} as well as nanostructures
\cite{ahlers:214411,ottaviano:242105,0957-4484-19-2-025707}. An ensemble of nanomagnets exhibits rich magnetic properties, with large
technological impact. Temperature-dependent memory effects and slow magnetic relaxation have been observed in a GaAs:Mn system containing
Mn-rich clusters \cite{wang:202503}. Despite numerous publications on the structural and magnetic properties of \mg~nanomagnets embedded in Ge
\cite{ahlers:214411,bihler:112506,ottaviano:242105,holy:144401,wang:101913,lechner:023102}, the information of their dynamic magnetization is
lacking. Jaeger \emph{et al.} reported a spin-glass-like behavior in Ge:Mn below 15 K \cite{jaeger:045330}. The effect was attributed to the
interaction between Mn-rich nanoclusters, and believed not to be related to \mg~precipitates. In this paper, we show that crystalline
\mg~nanomagnets are formed inside the Mn-diluted Ge matrix using Mn ion implantation. The Ge:\mg~system reveals pronounced temperature-dependent
memory effects and slow magnetic relaxation slightly below the superparamagnetic blocking temperature of \mg~nanomagnets that is much higher
than the blocking temperature of Mn-rich nanoclusters \cite{jaeger:045330}.

P-type doped Ge(100) single crystal wafers were implanted with 100 keV Mn ions to a fluence of $1\times10^{16}$ cm$^{-2}$, which corresponds to
a peak concentration of 2\% Mn. The samples were held
at 300 $^{\circ}$C during implantation to avoid amorphization. Structural analysis was performed by synchrotron radiation x-ray diffraction (SR-XRD)
at the Rossendorf beamline (BM20) at the ESRF with an
x-ray wavelength of 0.154 nm. Magnetic properties were analyzed using a superconducting quantum interference device (SQUID) magnetometer
(Quantum Design MPMS) with the field along the sample surface.

The SR-XRD 2$\theta$-$\theta$ scan confirms the formation of \mg~nanomagnets. As shown in Figure 1, beside the main peaks from Ge(004) and
Ge(002), the diffraction peaks of \mg(111), (002), (310), (222) and (004) are clearly visible. Note that compared to the work by Ottaviano
\emph{et al.} \cite{ottaviano:242105} SR-XRD reveals more \mg~peaks even for a much smaller Mn-ion fluence due to the large flux of x-rays. The
dilution of Mn ions inside Ge has been evidenced by photoemission spectroscopy \cite{ottaviano:242105} and by Hall measurements
\cite{riss:241202}. Mn ion implantation results in p-type doping of Ge \cite{riss:241202}. The ferromagnetic properties measured in this sample
are due to the formation of \mg~nanocrystals. The inset (a) of Figure 2 shows the hysteresis loop measured at 5 K. Note that the saturation
magnetization is around 0.75 $\mu_B$/Mn. This value is much smaller than 2.74 $\mu_B$/Mn for bulk \mg~\cite{bulk_Mn5Ge3}, which results from two
facts: Mn ions only partially form \mg~nanocrystals \cite{morgunov:085205} and the magnetization of nanoparticles is slightly smaller than that of bulk \mg~\cite{padova:045203}.

In order to study the time-dependent magnetization of \mg~nanomagnets, we performed history-dependent magnetic memory measurements using a
cooling and waiting protocol suggested by Sun \etal~ \cite{PhysRevLett.91.167206}. We cooled the sample at 50 Oe and recorded the magnetization
during cooling, but temporarily stopped at 200 K, 150 K, 100 K, 50 K, and 20 K for a waiting period of 2 hours. During waiting, the field was
set to zero. After the stop, the 50 Oe field was re-applied and cooling and measuring were resumed. The temporary stops resulted in a steplike
M(T) curve (solid black line) in Figure 2. After reaching 4 K, the sample was heated back in the same field, and the magnetization was recorded
again (dotted blue line). During this heating the M(T) curve also has a steplike behavior at the stop temperatures, then recovers the previous
M(T) curve measured during cooling, \emph{i.e.} the system remembers its thermal history. The steplike feature in the temperature-dependent
magnetization is a result of magnetic relaxation at the stopping points \cite{PhysRevLett.91.167206}. The observed memory effect and magnetic relaxation clearly demonstrate that
\mg~nanomagnets embedded inside a Ge matrix behave like a spin-glass \cite{jaeger:045330,wang:053912,PhysRevLett.91.167206}. We also measured the
thermo-remanent magnetization (TRM) as a function of time below and slightly above the blocking temperature (around 170 K) of \mg~nanomagnets.
TRM is measured by cooling the sample in an applied field of 50 Oe from 300 K to some final temperatures, decreasing the field to zero and
observing the decaying remanent magnetization. If the superparamagnetic nanoparticles undergo collective behavior due to direct dipole-dipole
interaction or particle size distribution, a stretched exponential form is expected \cite{jaeger:045330,wang:053912}:
\begin{equation}\label{relaxation_stretched}
    M_{r}(t)=M_0+M_{r}e^{-(t/\tau)^b},
\end{equation} where $\tau$ is the relaxation time, $b$ affects the relaxation of the glassy component, $M_r$ is the amplitude of the glassy component,
and $M_0$ is a time-independent constant term. Inset (b) of Figure 2 shows exemplarily the TRM time-decays at 100 K. It can be well fitted by
Eq. (\ref{relaxation_stretched}). The fitted parameters of $\tau$ at different temperatures are shown in Figure 4(c). Above the blocking
temperature, $\tau$ is comparable with the measurement time of around 100 seconds. The parameter $b$ is around 0.3, a typical value for a
spin-glass system \cite{jaeger:045330} and does not significantly change with temperature.

Figure 3 shows the magnetic relaxation after field cooling but with temporal temperature change. The sample was cooled in a field of 50 Oe from room
temperature to 100 K. Then the field was set to zero and the magnetization was recorded as a function of time. After a period of time $t_1$, the
sample was quenched to 90 K in a field of 50 Oe, and its magnetization was again recorded after setting the field to zero for a period of
time $t_2$ (temporal cooling). Finally, the temperature was increased back to 100 K in the field and magnetization was recorded after setting
the field to zero for a period of time $t_3$. For the temporal heating protocol, the sample was heated to 110 K during the period of time $t_2$.
The relaxation curve during \emph{t}$_3$ is the continuation of the relaxation during \emph{t}$_1$ after temporary cooling, but not after
temporary heating.

The memory effect as well as the relaxation phenomena observed support the view of a spin-glass like phase phase as described in detail in Ref.
\cite{PhysRevLett.91.167206}. Indeed, our observation is similar to the GaAs:Mn system consisting of Mn-rich clusters \cite{wang:202503} and to
permalloy nanoparticles embedded inside SiO$_2$ \cite{PhysRevLett.91.167206}. The origin of the spin-glass behavior is explained by considering
the inter-particle interaction. Although the direct dipole-dipole exchange is weak, Ruderman-Kittel-Kasuya-Yoshida (RKKY) interaction is
expected in the case of a Mn diluted GaAs or Ge host matrix which becomes highly p-type conducting due to Mn doping
\cite{wang:202503,jaeger:045330}. Within this frame, a memory effect is expected to be more pronounced when increasing the Mn concentration,
which results in a large density of nanomagnets. We also measured a sample with a much larger Mn fluence (5$\times$10$^{16}$ cm$^{-2}$) using
the same protocol, however, no memory effect was observable. Another explanation is a broad distribution of blocking temperatures originating
from the distribution of the particle size. The spin flip time for magnetic particles depends exponentially on the particle size. Therefore,
even a small distribution of the particle size could give a broad range of relaxation times \cite{PhysRevLett.91.167206,PhysRevB.72.014445}.

In order to corroborate this idea, we plot the relevant quantities vs temperature in Figure 4. The temperature dependent magnetization after
zero field cooling/field cooling (ZFC/FC) is shown in Figure 4(a). The broad peak in the ZFC curve is a direct reflection of the size
distribution of \mg~nanocrystals. The temperature-dependent magnetic remanence is also plotted in Figure 4(a). Below 15 K, the remanent
magnetization is increased steeply with decreasing temperature, which results from Mn-diluted Ge, a ferromagnetic phase with a \emph{T}$_C$
between 10-16 K \cite{bihler:112506,jaeger:045330,morgunov:085205}. Between 15 to 50 K, the remanence is weakly temperature-dependent: most of
the \mg~nanomagnets are large enough to overcome thermal fluctuation. In this temperature range, the remanence as a function of temperature
roughly obeys Bloch's \emph{T}$^{3/2}$ law. Above 50 K, the remanence drops more quickly with increasing temperature. Some small particles start
to flip due to thermal fluctuation. The gradient of the remanence can be used to estimate the number of particles changing from magnetically active (ferromagnetic)
to inactive (superparamagnetic) state. Strictly speaking, here we neglect the
gradient of saturated magnetization ($dM_{sat}/dT$), but it is reasonable since $dM_{sat}/dT$ is almost constant from 50 K to 200 K (not shown).
Figure 4(b) shows the differential of remanence versus temperature. Between 50 K and 150 K, a broad minimum with a dip at 135 K is observed.
The fast decay of the magnetic remanence indicates that a large number of particles have their blocking temperatures between 50 K to 150 K. In
this same temperature range, a peak occurs in the temperature dependent relaxation time ($\tau$) and the memory strength ($\Delta$\emph{M}) [Figure
4(c)]. $\Delta$\emph{M} is the change of magnetization before and after stopping as shown in Figure 2. Therefore, our observation strongly supports
that a size distribution of magnetic nanoparticles, rather than an inter-particle interaction, induces the striking memory effect. We have to
point out that the ZFC curve (not shown) for the sample with a large Mn fluence of 5$\times$10$^{16}$ cm$^{-2}$ reveals a sharper peak at higher
temperature due to a narrower size distribution and no memory effect compared with the sample presented here. However, our finding does not
exclude the explanation in Refs. \onlinecite{wang:202503,jaeger:045330}, in which the authors investigated Mn-rich nanoclusters (not necessarily
to be crystalline precipitates).

In conclusion, we have synthesized crystalline \mg~nanomagnets embedded inside a Mn diluted Ge matrix using ion implantation. A striking memory
effect and slow magnetic relaxation are observed at temperatures near the superparamagnetic blocking temperature. Our findings corroborate that
the spin-glass-like features are caused by the size distribution of \mg~nanomagnets, rather than by the inter-particle interaction through the
Mn-diluted matrix. A recent theoretical work on the spin-glass behavior in GaAs:Mn consisting of Mn-rich nanoclusters by Chang \emph{et al.}
also supports this conclusion \cite{chang:212106}.

The author (S.Z.) acknowledges the financial support from the Bundesministerium f\"{u}r Bildung und Forschung (FKZ13N10144) and A.S. wants to
thank the Deutsche Forschungsgemeinschaft (DFG) (PO1275/2-1, 'SEMAN').


\clearpage

\begin{figure} \center
\includegraphics[scale=0.8]{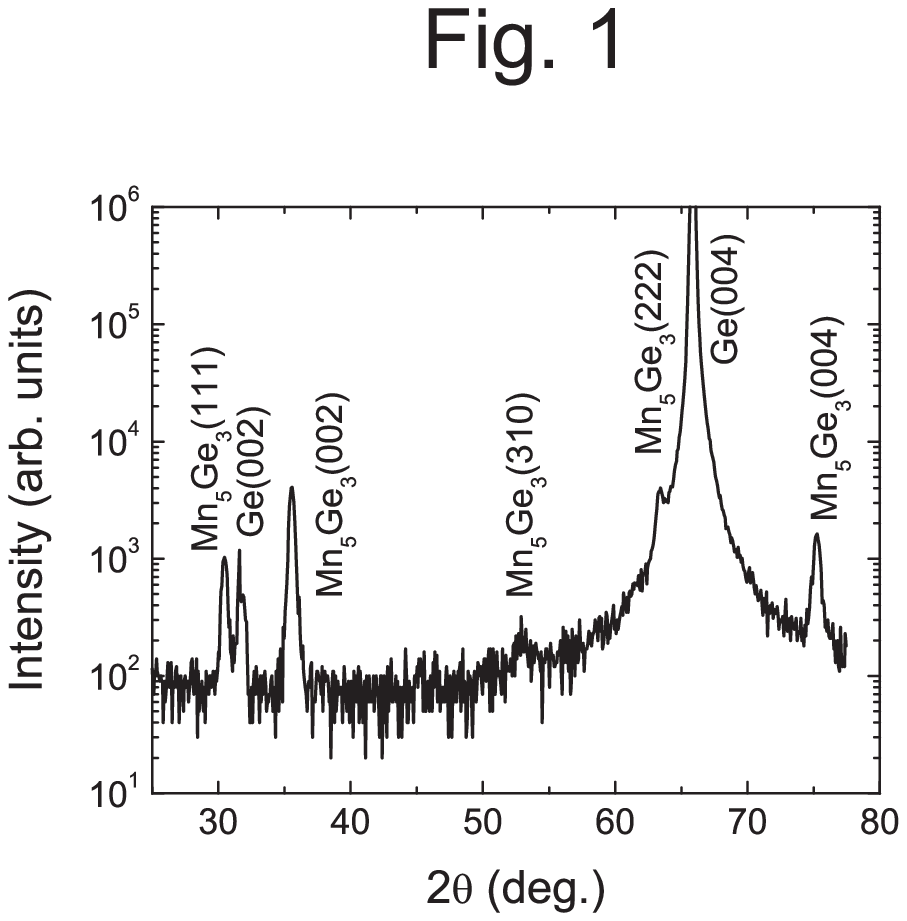}
\end{figure}
Fig. 1 XRD 2$\theta$-$\theta$ scan revealing the formation of \mg~nanomagnets. Beside the main peaks from Ge(004) and Ge(002), the diffraction
peaks of \mg(111), (002), (310), (222) and (004) are clearly visible.

\begin{figure} \center
\includegraphics[scale=1]{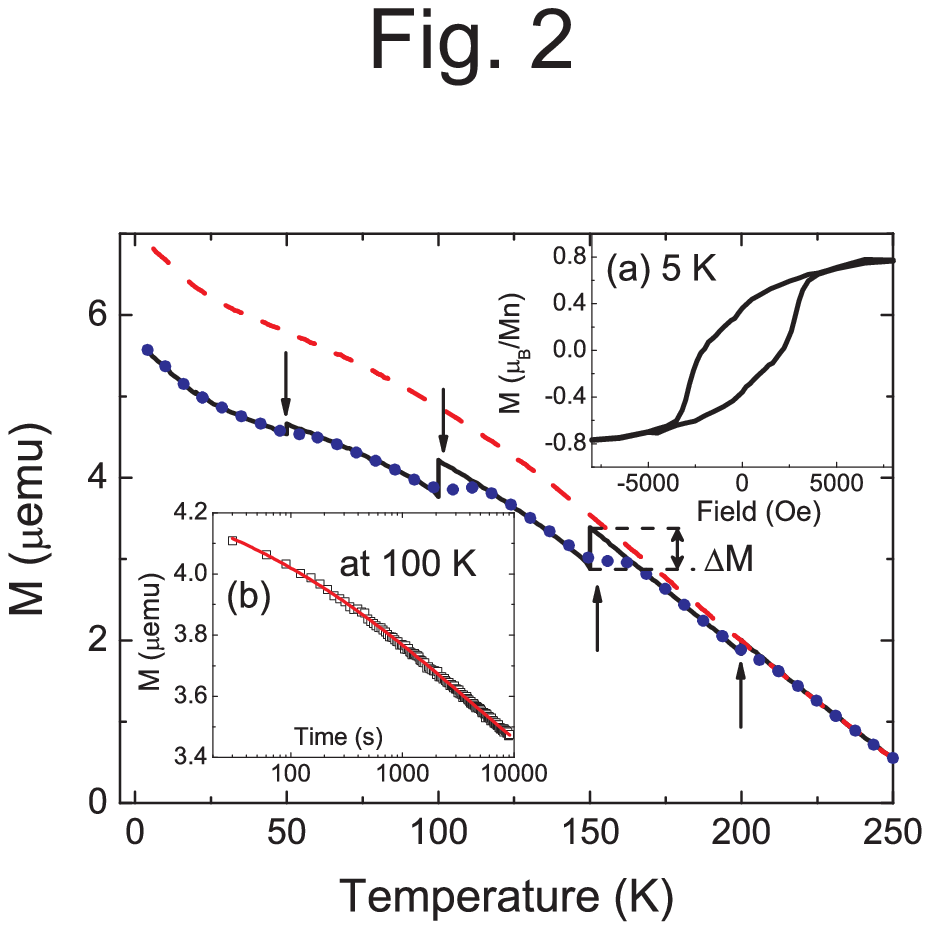}
\end{figure}
Fig. 2 Temperature dependent memory effect in the dc magnetization. The dashed line (red) is measured during cooling in 50 Oe at a cooling rate
of 1 K per minute, while the solid line (black) is measured in 50 Oe with the same cooling rate but with a stop of 2 hours at 200 K, 150 K, 100
K, 50 K and 20 K. The field is cut off during stop. The magnetization change (${\Delta}M$) before and after stopping is observed at 150 K. The
dotted line (blue) is measured with continuous heating at the same rate after the previous cooling protocol. Inset (a): Hysteresis loop measured
at 5 K. Inset (b): time dependent remanent magnetization measured after cooling from 300 K to 100 K with a field of 50 Oe. Scattered symbols are
experimental data and the solid line (blue) is a fitting using the stretched-exponential function [Eq. (\ref{relaxation_stretched})].

\begin{figure} \center
\includegraphics[scale=0.8]{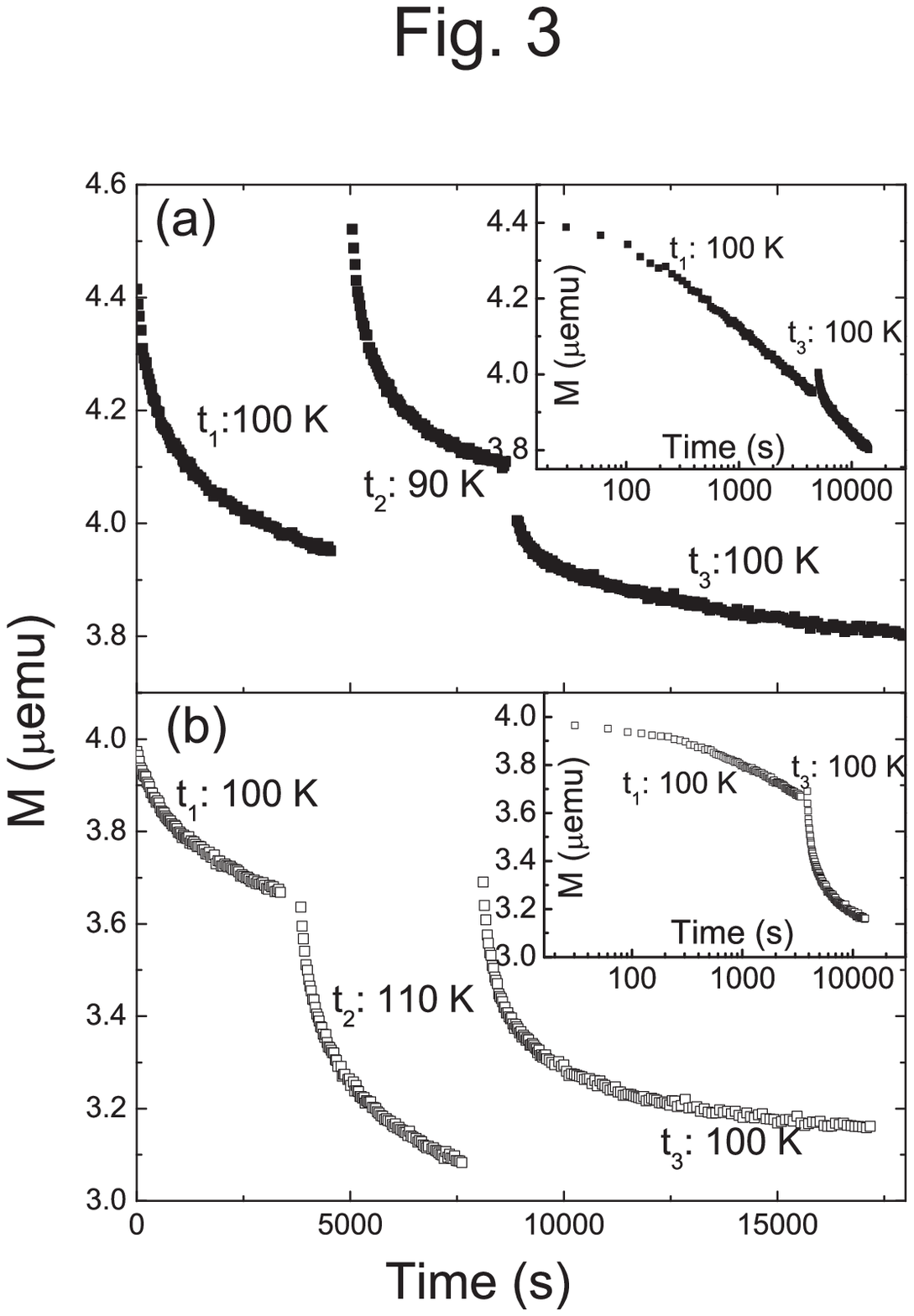}
\end{figure}
Fig. 3 Magnetic relaxation after field cooling (a) with temporary cooling at $T_0+\Delta{T}=100 K-10 K$ and (b) with temporary heating at
$T_0+\Delta{T}=100 K+10 K$. The insets plot the same data vs the total time spent at 100 K. The relaxation curve during \emph{t}$_3$ is the
continuation of that during \emph{t}$_1$ after temporary cooling, but not after temporary heating. After temporary heating, no memory effect is
observable.

\begin{figure} \center
\includegraphics[scale=0.8]{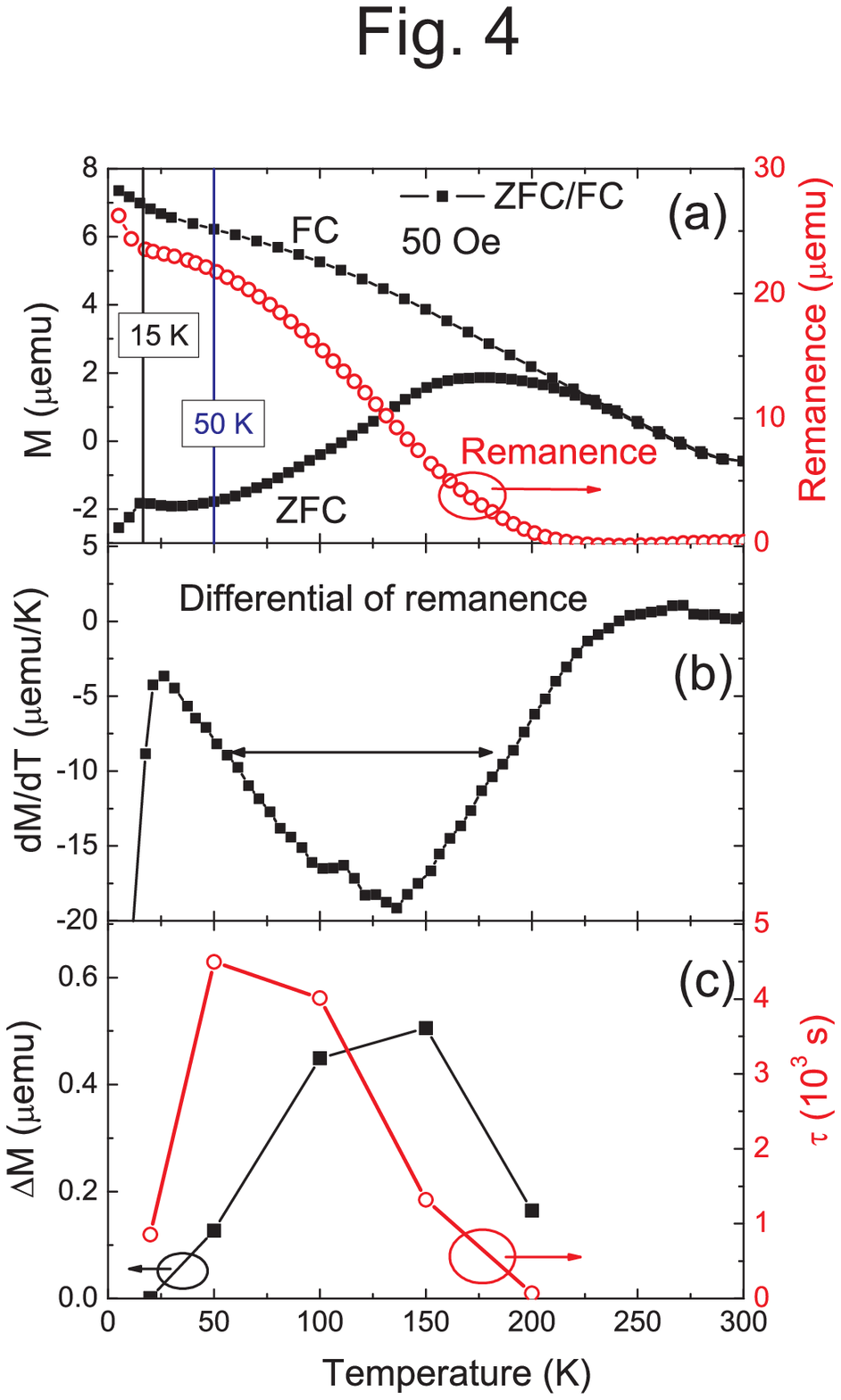}
\end{figure}
Fig. 4 Comparison between different magnetization quantities depending on temperature (a) ZFC/FC curves using a field of 50 Oe. The magnetic
remanence is also shown. To measure the magnetic remanence, a field of 8000 Oe was applied to saturate the magnetization at 5 K and the remanence was measured during warming in a zero field.
(b) The differential of magnetic remanence versus temperature revealing a broad minimum indicated by the arrow. (c) The
magnetization change before and after stopping (${\Delta}M$ as shown in Figure 2) and the relaxation time $\tau$ obtained by fitting using Eq.
(\ref{relaxation_stretched}).

\end{document}